\documentclass[a4paper,twoside,11pt]{article}
\usepackage{amsfonts}
\usepackage{amssymb}\usepackage{amsmath}
\usepackage{graphicx}
\def\be{\begin{equation}}
\def\ee{\end{equation}}
\def\ba{\begin{array}}
\def\ea{\end{array}}
\def\bc{\begin{center}}
\def\ec{\end{center}}

\def\ZZ{\rm {{\rm Z}\kern-.48em{\rm Z}}}
\def\RR{\rm \hbox{I\kern-.2em\hbox{R}}}
\def\CC{\rm \hbox{C\kern -.5em{\raise .32ex
\hbox{$\scriptscriptstyle |$}}\kern - .22em{\raise .6ex
\hbox{$\scriptscriptstyle |$}}\kern
.4em}}
\def\CC{ {\mathbf C} }
\def\RR{ {\mathbf R} }
\def\ZZ{ {\mathbf Z} }

\newtheorem{Tm}{Theorem}[section]
\newtheorem{Df}[Tm]{Definition}

\newtheorem{Lm}[Tm]{Lemma}

  \newtheorem{exam}{Example}[section]
 \numberwithin{equation}{section}
\newcommand{\double}{\baselineskip 1.24 \baselineskip}

\title{ Adaptive sampling of time-space signals in a reproducing kernel subspace of mixed Lebesgue space
}
\author{{\small  Yingchun Jiang$^{1}$, \ \ Wenchang Sun$^{2} $}\\
{\small
1. School of Mathematics and Computational Science,}\\{\small
Guilin University of Electronic Technology, Guilin 541004, China}
\\{\small
2. School of Mathematical Sciences and LPMC, Nankai University, Tianjin 300071, China}}

\begin{document}
\date{}
\maketitle \double
\textbf{Abstract:}\ The Mixed Lebesgue space is a suitable tool for modelling and measuring signals living in time-space domains. And sampling in such spaces  plays an important role for processing high-dimensional time-varying signals. In this paper, we first define  reproducing kernel subspaces of mixed Lebesgue spaces. Then, we study the frame properties and show that the  reproducing kernel subspace has finite rate of innovation. Finally, we propose a semi-adaptive sampling scheme for time-space signals in a reproducing kernel subspace, where the sampling in time domain is conducted by a time encoding machine. Two kinds of timing  sampling methods are considered and the corresponding iterative approximation algorithms with exponential convergence are given.

\textbf{Keywords:} time-space signals; reproducing kernel subspace; mixed Lebesgue space; time encoding machine; semi-adaptive sampling

 {\bf MR(2000) Subject Classification:}  94A20, 46E22.

\section{Introduction }

\ \ \ \ \ \ In  practice, some signals are time-varying and the mixed Lebesgue space is a suitable tool for measuring such signals. Mixed Lebesgue spaces arise for considering functions that depend on independent quantities with different properties, which were first described in detail in \cite{bp} and were furtherly studied in \cite{bc,bz,f,frt}. In fact, the flexibility for the separate integrability of each variable had been generally applied in the study of time-based partial differential equations.

The mixed Lebesgue space $L^{p,q}(R^{d+1})$ consists of all measurable functions $f=f(x,y)$ defined on $R\times R^{d}$ such that
\begin{equation}
||f||_{L^{p,q}}=\big\|\|f(x,y)\|_{L^q_y(R^d)}\big\|_{L^p_x(R)}<\infty, 1\leq p,q\leq\infty.
\end{equation}
The corresponding sequence spaces are defined by
\begin{equation}
\ell^{p,q}(Z^{d+1})=\Big\{c:||c||_{\ell^{p,q}}=\big\|\|c(k_1,k_2)\|_{\ell^q_{k_2}}\big\|_{\ell^p_{k_1}}<\infty\Big\},  1\leq p,q\leq\infty.
\end{equation}
 Obviously, $L^{p,p}(R^{d+1})=L^{p}(R^{d+1})$ and $\ell^{p,p}(Z^{d+1})=\ell^{p}(Z^{d+1})$.

Sampling is an important task  in signal and image processing. There are many results for sampling and reconstruction of various signals, such as bandlimited signals \cite{s49,s2}, signals in shift-invariant spaces \cite{ag,lw,u50}, signals with finite rate of innovation \cite{s06} and signals in a reproducing kernel subspace \cite {hns,j16,ns,x10}. Sampling for signals living in a mixed Lebesgue space is useful for processing time-based signals. In fact, Sampling of band-limited signals in mixed Lebesgue spaces was studied in \cite{tw,w10}. Recently, nonuniform sampling in shift-invariant subspaces of $L^{p,q}(R^{d+1})$ was discussed in \cite{lllz}.

In this paper, we study the sampling and reconstruction of signals in a reproducing kernel subspace of  $L^{p,q}(R^{d+1})$.
The classical sampling sets are not adaptive to signals and the sampling process is linear. Recently, a time sampling approach called time encoding machine (TEM) has received attention \cite{TEM04,TEM104,TEM06,TEM05}, which is inspired by the neurons models.  Instead of recording the value of a signal $f(t)$ at a preset time instant, one records the
time at which the signal takes on a preset value. So it is a signal-dependent and nonlinear sampling mechanism. It is more practical in practice,  due to its simplicity and low-cost for sampling. A time encoding machine maps amplitude information of a signal into the timing domain, which was first introduced by Lazar and T\'{o}th in \cite{TEM04} for the special case of bandlimited signals and was extended to  $L^2$-shift-invariant subspaces \cite{gv} and more general framework of weighted reproducing kernel subspaces \cite{j16}.

Since the signals $f(x,y)$ in our setting live in the time-space domains, we assume that some sampling devices with function of time encoding are located at  $\Gamma=\{y_j: j\in J\}\subset R^d$. Each device living on $y_j,j\in J$ first takes samples $f(x,y_j)$ in space domain, and then  produces samples for time domain by time encoding machines. $\Gamma$  is supposed to be relatively-separated, that is,
\begin{equation*}
B_{\Gamma}(\delta')=\sup \limits_{y\in R^d}\sum \limits_{j\in J}\chi_{B(y_j,\delta')}(y)<\infty
\end{equation*}
for some  $\delta'>0$. Furthermore,  $\delta'>0$ is said to be the gap of $\Gamma$ if
\begin{equation*}
A_{\Gamma}(\delta')=\inf \limits_{y\in R^d}\sum \limits_{j\in J}\chi_{B(y_j,\delta')}(y)\geq 1.
\end{equation*}
Here, $J$ is a countable index set,  $B(y,\delta')$ are ball in $R^d$ with radius $\delta'$.

This paper is organized as follows. In section 2, we define the reproducing kernel subspaces of  mixed Lebesgue spaces $L^{p,q}(R^{d+1})$ and give a class of examples.  In
section 3, the frame properties of  reproducing kernel subspaces are studied. Section 4 is devoted to presenting two kinds of time encoding machines and demonstrating the iterative reconstruction algorithms for recovering  signals in a reproducing kernel  subspace of mixed Lebesgue space.

\section{Reproducing kernel subspaces in $L^{p,q}(R^{d+1})$}

\ \ \ \ \ \ In general, the reconstruction of signals can not be solved efficiently without extra smooth information on signals. In this section, we define a reproducing kernel subspace of mixed Lebesgue space $L^{p,q}(R^{d+1})$ for modelling the time-space signals.

Suppose that $K$ is a function defined on $(R\times R^d)\times(R\times R^d)$, which satisfies
\begin{equation}\label{r1}
\|K\|_{\mathcal{W}}:=\big\|\|K(x,y;s,t)\|_{\mathcal{W}^0_{y,t}}\big\|_{\mathcal{W}^0_{x,s}}<\infty
\end{equation}
and
\begin{equation}\label{r2}
\lim \limits_{\delta\rightarrow 0}\|\omega_\delta(K)\|_{\mathcal{W}}=0.
\end{equation}
Here, $\omega_\delta(K)$ is the modulus of continuity defined by
\begin{equation*}
\omega_\delta(K)(x,y;s,t):=\sup \limits_{|(x',y')|\leq \delta,|(s',t')|\leq \delta}|K(x+x',y+y';s+s',t+t')-K(x,y;s,t)|.
\end{equation*}
For a function $K_0(x,y)$ defined on $R^n\times R^n$, the $\mathcal{W}^0$-norm is defined as
\begin{equation}\label{x1}
\|K_0\|_{\mathcal{W}^0}:=\max\Big\{\sup \limits_{x\in R^n}\|K_0(x,\cdot)\|_{L^1(R^n)},\sup \limits_{y\in R^n}\|K_0(\cdot,y)\|_{L^1(R^n)}\Big\}.
\end{equation}
Let $T$ be an idempotent ($T^2=T$) integral operator on $L^{p,q}(R^{d+1})$ with kernel $K$,
\begin{equation}\label{r3}
Tf(x,y):=\int_R\int_{R^d}K(x,y;s,t)f(s,t)dsdt,\ f\in L^{p,q}(R^{d+1}).
\end{equation}
Then $T$ is a bounded operator on $L^{p,q}(R^{d+1})$, which can be proved by the following two lemmas.
\begin{Lm}(Minkowski's \ inequality)
Let $1\leq p\leq \infty$. Suppose that $f(x,y)$ is a measurable function on $R^m\times R^n$\ $(m,n\in N)$. Then
$$\big\|\int_{R^n}|f(\cdot,y)|dy\big\|_{L^p(R^m)}\leq \int_{R^n}\|f(\cdot,y)\|_{L^p(R^m)}dy.$$
\end{Lm}
\begin{Lm}\cite {ns}
Let $T_0$ be an integral operator on $L^p(R^d)$ defined by
$$T_0f(x)=\int_{R^d}K_0(x,y)f(y)dy.$$
If the kernel $K_0$ satisfies $\|K_0\|_{\mathcal{W}^0}<\infty$,
then $\|T_0f\|_{L^p}\leq \|K_0\|_{\mathcal{W}^0}\|f\|_{L^p}.$
\end{Lm}

\begin{Lm}\label{L2.3}
Let $T$ be the  integral operator on $L^{p,q}(R^{d+1})$ defined in \eqref{r3}. If $K$ satisfies the condition \eqref{r1}, then
$$\|Tf\|_{L^{p,q}}\leq \|K\|_{\mathcal{W}}\|f\|_{L^{p,q}},\ f\in L^{p,q}(R^{d+1}).$$
\end{Lm}
{\bf Proof}\ It follows from Lemma 2.1 and Lemma 2.2 that
\begin{align*}
\|Tf\|_{L^{p,q}}&=\Big\|\int_R\big(\int_{R^d}K(x,y;s,t)f(s,t)dt\big)ds\Big\|_{L^{p,q}}\\
&=\Big\|\big\|\int_R\big(\int_{R^d}K(x,y;s,t)f(s,t)dt\big)ds\big\|_{L^q_y(R^d)}\Big\|_{L^p_x(R)}\\
&\leq \Big\|\int_R\big\|\int_{R^d}K(x,y;s,t)f(s,t)dt\big\|_{L^q_y(R^d)}ds\Big\|_{L^p_x(R)}\\
& \leq\Big\|\int_R\|K(x,y;s,t)\|_{\mathcal{W}^0_{y,t}}\|f(s,\cdot)\|_{L^q(R^d)}ds\Big\|_{L^p_x(R)}\\
&\leq \|K\|_{\mathcal{W}}\|f\|_{L^{p,q}}.
\end{align*}

The following main theorem of this section  show that the range space of the operator $T$ in \eqref{r3} is a reproducing kernel subspace of $L^{p,q}(R^{d+1})$ under suitable conditions for the kernel $K$.

\begin{Tm}\label{T21}
Let $V$ be the range space of the operator $T$, that is,
\begin{equation}\label{r26}
V=\{Tf:\ f\in L^{p,q}(R^{d+1})\}=\{f\in L^{p,q}(R^{d+1}):\ Tf=f\}.
\end{equation}
If the  kernel $K$ satisfies \eqref{r1} and \eqref{r2}, then
 \begin{enumerate}
\item[(i)]  $V$ is a reproducing kernel subspace of $L^{p,q}(R^{d+1})$, that is, for any $(x,y)\in R\times R^d$, there exists a constant $C_{x,y}>0$ such that
$$|f(x,y)|\leq C_{x,y}\|f\|_{L^{p,q}},\ f\in V;$$
\item[(ii)]  The kernel $K$ satisfies the "reproducing kernel property"
\begin{equation}\label{r12}
\int_R\int_{R^d}K(x,y;u,v)K(u,v;s,t)dudv=K(x,y;s,t),\ \forall\ (x,y),(s,t)\in R\times R^d;
\end{equation}
\item [(iii)] For any $(s,t)\in R\times R^d$, $K(\cdot,\cdot;s,t)\in V$.
\end{enumerate}
\end{Tm}

To prove this theorem, we need the following lemma.
\begin{Lm}\cite {bp}
Let $1\leq p,q\leq \infty$, $\frac{1}{p}+\frac{1}{p'}=1$ and $\frac{1}{q}+\frac{1}{q'}=1$. Then
\begin{equation*}
\|fg\|_{L^1}\leq \|f\|_{L^{p,q}}\|g\|_{L^{p',q'}}.
\end{equation*}
\end{Lm}
{\bf Proof of Theorem \ref{T21}} \ (i)\ For $f(x,y)\in V$, we have
\begin{align*}
|f(x,y)|&=|Tf(x,y)|\\
&=|\int_R\int_{R^d}K(x,y;s,t)f(s,t)dsdt|\\
&\leq \|f\|_{L^{p,q}}\|K(x,y;\cdot,\cdot)\|_{L^{p',q'}}.
\end{align*}
In the following, we estimate $\|K(x,y;\cdot,\cdot)\|_{L^{p',q'}}$. On the one hand,
\begin{align}\label{r5}
\|K(x,y;s,t)\|_{L^1_t}&=\int_{R^d}|K(x,y;s,t)|dt\nonumber\\
&\leq \sup \limits_{y\in R^d}\int_{R^d}|K(x,y;s,t)|dt\nonumber\\
&\leq \|K(x,y;s,t)\|_{\mathcal{W}^0_{y,t}}.
\end{align}
On the other hand, for  $(x,y)\in R\times R^d$, by the definition of the modulus of continuity, one has
\begin{align}\label{r4}
|K(x,y;s,t)|&\leq \delta^{-(d+1)}\int_{k_1\delta+[-\delta/2,\delta/2]}\int_{k_2\delta+[-\delta/2,\delta/2]^d}\Big(|K(x,y;u,v)|+\omega_{\sqrt{d+1}\delta}(K)(x,y;u,v)\Big)dudv\nonumber\\
&=:\delta^{-(d+1)}K_1(x,y;s,t),
\end{align}
where $s\in k_1\delta+[-\delta/2,\delta/2]$, $t\in k_2\delta+[-\delta/2,\delta/2]^d$, $k_1\in Z, k_2\in Z^d$.

Note that for $s\in k_1\delta+[-\delta/2,\delta/2]$, one has
\begin{align}
\|K_1(x,y;s,t)\|_{L^\infty_t}&\leq \int_{k_1\delta+[-\delta/2,\delta/2]}\Big(\|K(x,y;u,t)\|_{\mathcal{W}^0_{y,t}}+\|\omega_{\sqrt{d+1}\delta}(K)(x,y;u,t)
\|_{\mathcal{W}^0_{y,t}}\Big)du\nonumber\\
&=:K_2(x,s).
\end{align}
This together with \eqref{r4} shows that
\begin{equation}\label{r6}
\|K(x,y;s,t)\|_{L^\infty_t}\leq \delta^{-(d+1)}K_2(x,s).
\end{equation}
Now, it follows from \eqref{r5} and \eqref{r6} that
\begin{align}\label{r10}
\|K(x,y;s,t)\|_{L^{q'}_t}&\leq \|K(x,y;s,t)\|_{L^{1}_t}^{\frac{1}{q'}}\|K(x,y;s,t)\|_{L^{\infty}_t}^{1-\frac{1}{q'}}\nonumber\\
&\leq \|K(x,y;s,t)\|_{\mathcal{W}^0_{y,t}}^{\frac{1}{q'}}\delta^{-(d+1)(1-\frac{1}{q'})}\big(K_2(x;s)\big)^{1-\frac{1}{q'}}\nonumber\\
&=:\delta^{-(d+1)(1-\frac{1}{q'})}K_3(x;s).
\end{align}
Next, we  estimate $\|K_3(x;s)\|_{L^1_s}$ and $\|K_3(x;s)\|_{L^\infty_s}$. In fact,
\begin{align}\label{r11}
\|K_3(x;s)\|_{L^1_s}&\leq \Big(\int_R\|K(x,y;s,t)\|_{\mathcal{W}^0_{y,t}}ds\Big)^{\frac{1}{q'}}\Big(\int_RK_2(x;s)ds\Big)^{1-\frac{1}{q'}}\nonumber\\
&\leq \|K\|_{\mathcal{W}}^{\frac{1}{q'}}\Big(\sum \limits_{k_1\in Z}\int_{k_1\delta+[-\delta/2,\delta/2]}K_2(x;s)ds\Big)^{1-\frac{1}{q'}}\nonumber\\
&=\|K\|_{\mathcal{W}}^{\frac{1}{q'}}\Big(\sum \limits_{k_1\in Z}\int_{k_1\delta+[-\delta/2,\delta/2]}\Big[\int_{k_1\delta+[-\delta/2,\delta/2]}\Big(\|K(x,y;u,t)\|_{\mathcal{W}^0_{y,t}}\nonumber\\
&+\|\omega_{\sqrt{d+1}\delta}(K)(x,y;u,t)
\|_{\mathcal{W}^0_{y,t}}\Big)du\Big]ds\Big)^{1-\frac{1}{q'}}\nonumber\\
&=\|K\|_{\mathcal{W}}^{\frac{1}{q'}}\Big(\delta\int_R\Big(\|K(x,y;u,t)\|_{\mathcal{W}^0_{y,t}}+\|\omega_{\sqrt{d+1}\delta}(K)(x,y;u,t)
\|_{\mathcal{W}^0_{y,t}}\Big)du\Big)^{1-\frac{1}{q'}}\nonumber\\
&\leq \|K\|_{\mathcal{W}}^{\frac{1}{q'}}\delta^{1-\frac{1}{q'}}\Big(\|K\|_{\mathcal{W}}+\|\omega_{\sqrt{d+1}\delta}(K)\|_{\mathcal{W}}\Big)^{1-\frac{1}{q'}}.
\end{align}
Moreover, we have
\begin{align}\label{r7}
\|K_3(x;s)\|_{L^\infty_s}&\leq \Big(\sup \limits_{s\in R}\|K(x,y;s,t)\|_{\mathcal{W}^0_{y,t}}\Big)^{\frac{1}{q'}}\Big(\sup \limits_{s\in R}K_2(x;s)\Big)^{1-\frac{1}{q'}}\nonumber\\
&\leq \Big(\|K\|_{\mathcal{W}}+\|\omega_{\sqrt{d+1}\delta}(K)\|_{\mathcal{W}}\Big)^{1-\frac{1}{q'}}\Big(\sup \limits_{s\in R}\|K(x,y;s,t)\|_{\mathcal{W}^0_{y,t}}\Big)^{\frac{1}{q'}}.
\end{align}
Furthermore, for $s\in k_1\delta+[-\delta/2,\delta/2]$, if follows from \eqref{r4} that
\begin{align*}
\int_R|K(x,y;s,t)|dt&\leq \delta^{-(d+1)}\sum \limits_{k_2\in Z^d}\int_{k_2\delta+[-\delta/2,\delta/2]^d}\int_{k_1\delta+[-\delta/2,\delta/2]}\int_{k_2\delta+[-\delta/2,\delta/2]^d}\nonumber\\
&\Big(|K(x,y;u,v)|+\omega_{\sqrt{d+1}\delta}(K)(x,y;u,v)\Big)dudvdt\nonumber\\
&=\delta^{-1}\int_{k_1\delta+[-\delta/2,\delta/2]}\int_{R^d}\Big(|K(x,y;u,v)|+\omega_{\sqrt{d+1}\delta}(K)(x,y;u,v)\Big)dudv\nonumber\\
&\leq \delta^{-1}\int_{k_1\delta+[-\delta/2,\delta/2]}\Big(\|K(x,y;u,t)\|_{\mathcal{W}^0_{y,t}}+\|\omega_{\sqrt{d+1}\delta}(K)(x,y;u,t)
\|_{\mathcal{W}^0_{y,t}}\Big)du\nonumber\\
&\leq \delta^{-1}\Big(\|K\|_{\mathcal{W}}+\|\omega_{\sqrt{d+1}\delta}(K)\|_{\mathcal{W}}\Big).
\end{align*}
Therefore, we have
\begin{equation}\label{r8}
\sup \limits_{s\in R}\sup \limits_{y\in R^d}\int_R|K(x,y;s,t)|dt\leq \delta^{-1}\Big(\|K\|_{\mathcal{W}}+\|\omega_{\sqrt{d+1}\delta}(K)\|_{\mathcal{W}}\Big).
\end{equation}
Similarly, we can obtain
\begin{equation}\label{r9}
\sup \limits_{s\in R}\sup \limits_{t\in R^d}\int_R|K(x,y;s,t)|dy\leq \delta^{-1}\Big(\|K\|_{\mathcal{W}}+\|\omega_{\sqrt{d+1}\delta}(K)\|_{\mathcal{W}}\Big)
\end{equation} from the similar estimation
$$|K(x,y;s,t)|\leq \delta^{-(d+1)}\int_{k_1\delta+[-\delta/2,\delta/2]}\int_{k_2\delta+[-\delta/2,\delta/2]^d}\Big(|K(x,v;u,t)|+\omega_{\sqrt{d+1}\delta}(K)(x,v;u,t)\Big)dudv$$
as \eqref{r4}, where $s\in k_1\delta+[-\delta/2,\delta/2]$ and $y\in k_2\delta+[-\delta/2,\delta/2]^d$. Thus, it follows from \eqref{r7}, \eqref{r8} and \eqref{r9} that
\begin{equation}
\|K_3(x;s)\|_{L^\infty_s}\leq \delta^{-\frac{1}{q'}}\Big(\|K\|_{\mathcal{W}}+\|\omega_{\sqrt{d+1}\delta}(K)\|_{\mathcal{W}}\Big).
\end{equation}
This together with \eqref{r10} and \eqref{r11} shows that
\begin{align}\label{r13}
\|K(x,y;\cdot,\cdot)\|_{L^{p',q'}}&\leq \delta^{-(d+1)(1-\frac{1}{q'})}\|K_3(x;s)\|_{L^{p'}_s}\nonumber\\
&\leq \delta^{-(d+1)(1-\frac{1}{q'})}\|K_3(x;s)\|_{L^{1}_s}^{\frac{1}{p'}}\|K_3(x;s)\|_{L^{\infty}_s}^{1-\frac{1}{p'}}\nonumber\\
&\leq\delta^{-d(1-\frac{1}{q'})-(1-\frac{1}{p'})}\|K\|_{\mathcal{W}}^{\frac{1}{p'q'}} \Big(\|K\|_{\mathcal{W}}+\|\omega_{\sqrt{d+1}\delta}(K)\|_{\mathcal{W}}\Big)^{1-\frac{1}{p'q'}}.
\end{align}

\noindent(ii)\ Let $A(x,y;s,t)=:\int_R\int_{R^d}K(x,y;u,v)K(u,v;s,t)dudv.$ For fixed $x,s\in R$,
\begin{align*}
\sup \limits_{y\in R^d}\int_{R^d}|A(x,y;s,t)|dt&\leq \sup \limits_{y\in R^d}\int_R\int_{R^d}|K(x,y;u,v)|\Big[\int_{R^d}|K(u,v;s,t)|dt\Big]dudv\\
&\leq \sup \limits_{y\in R^d}\int_R\int_{R^d}|K(x,y;u,v)|\|K(u,v;s,t)\|_{\mathcal{W}^0_{v,t}}dudv\\
& \leq \int_R\|K(u,v;s,t)\|_{\mathcal{W}^0_{v,t}}\sup \limits_{y\in R^d}\Big[\int_{R^d}|K(x,y;u,v)|dv\Big]du\\
&\leq \int_R\|K(x,y;u,v)\|_{\mathcal{W}^0_{y,v}}\|K(u,v;s,t)\|_{\mathcal{W}^0_{v,t}}du.
\end{align*}
Similarly, we can obtain
$$\sup \limits_{t\in R^d}\int_{R^d}|A(x,y;s,t)|dy\leq\int_R\|K(x,y;u,v)\|_{\mathcal{W}^0_{y,v}}\|K(u,v;s,t)\|_{\mathcal{W}^0_{v,t}}du.$$
Therefore, one has $$\|A(x,y;s,t)\|_{\mathcal{W}^0_{y,t}}\leq\int_R\|K(x,y;u,v)\|_{\mathcal{W}^0_{y,v}}\|K(u,v;s,t)\|_{\mathcal{W}^0_{v,t}}du=:A_1(x;s).$$
Furthermore,
\begin{align*}
\sup \limits_{x\in R}\int_{R}|A_1(x;s)|ds&= \sup \limits_{x\in R}\int_R\|K(x,y;u,v)\|_{\mathcal{W}^0_{y,v}}\Big[\int_{R}\|K(u,v;s,t)\|_{\mathcal{W}^0_{v,t}}ds\Big]du\\
&\leq \sup \limits_{x\in R}\int_R\|K(x,y;u,v)\|_{\mathcal{W}^0_{y,v}}\|K\|_{\mathcal{W}}du\\
&\leq\|K\|_{\mathcal{W}}^2.
\end{align*}
Similarly, $\sup \limits_{s\in R}\int_{R}|A_1(x;s)|dx\leq \|K\|_{\mathcal{W}}^2$.  Moreover, $\|A\|_{\mathcal{W}}\leq \|K\|_{\mathcal{W}}^2.$ Therefore, the kernel
$$B(x,y;s,t)=:A(x,y;s,t)-K(x,y;s,t)$$
of the linear operator $T^2-T$ satisfies
$$\|B\|_{\mathcal{W}}\leq \|A\|_{\mathcal{W}}+\|K\|_{\mathcal{W}}\leq \|K\|_{\mathcal{W}}(1+\|K\|_{\mathcal{W}})<\infty.$$
Finally, \eqref{r12} follows from $T^2=T$.

\noindent(iii)\ Note that \eqref{r12} holds, we only need to verify that $K(\cdot,\cdot;s,t)\in L^{p,q}$. In fact, by the similar method for proving \eqref{r13}, we can obtain
\begin{equation}\label{r17}
\|K(\cdot,\cdot;s,t)\|_{L^{p,q}}\leq\delta^{-d(1-\frac{1}{q})-(1-\frac{1}{p})}\|K\|_{\mathcal{W}}^{\frac{1}{pq}} \Big(\|K\|_{\mathcal{W}}+\|\omega_{\sqrt{d+1}\delta}(K)\|_{\mathcal{W}}\Big)^{1-\frac{1}{pq}}
\end{equation}
for $\delta>0$.

In the rest of this section, we  give an example of the reproducing kernel subspace $V$ of $L^{p,q}(R^{d+1})$. We say that a measurable function $f(x,y)$ defined on $R\times R^{d}$ belongs to the
Wiener amalgam space $W(L^1)(R^{d+1})$  if it satisfies
$$\|f\|_{W(L^1)}=\sum \limits_{k_1\in Z}\sum \limits_{k_2\in Z^d}\sup \limits_{(x,y)\in [0,1]\times[0,1]^{d}}|f(x+k_1,y+k_2)|<\infty.$$
We refer more details about Wiener amalgam spaces and their applications to \cite{fei90}.
\begin{Lm}\cite{ag,lllz} \label{L2.6}
Let $\varphi(x,y)\in W(L^1)(R^{d+1})$ be continuous and satisfy $$0<m\leq \sum \limits_{k\in Z^{d+1}}|\widehat{\varphi}(\xi+2k\pi)|^2\leq M<\infty.$$
Then the following results hold:

\noindent(i)\ The dual generator $\widetilde{\varphi}(x,y)$ is also in $W(L^1)(R^{d+1})$ and
\begin{equation}\label{r30}
\widetilde{\varphi}(x,y)=\sum \limits_{k_1\in Z}\sum \limits_{k_2\in Z^d}b(k_1,k_2)\varphi(x-k_1,y-k_2),
\end{equation}
where $b\in \ell^1(Z^{d+1})$.

\noindent(ii)\ The modulus of continuity
$$\omega_{\delta}(\varphi)(x,y)=\sup \limits_{|(x',y')|\leq \delta}|\varphi(x+x',y+y')-\varphi(x,y)|$$
satisfies $\lim \limits_{\delta\rightarrow 0}\|\omega_\delta(\varphi)\|_{W(L^1)}=0$.

\noindent(iii)\ Let $1\leq p,q\leq \infty$. The shift-invariant space
\begin{equation}\label{r27}
V_{p,q}(\varphi)=\Big\{\sum \limits_{k_1\in Z}\sum \limits_{k_2\in Z^d}c(k_1,k_2)\varphi(x-k_1,y-k_2):\ \{c(k_1,k_2)\}_{k_1\in Z, k_2\in Z^d}\in \ell^{p,q}(Z^{d+1})\Big\}
\end{equation}
is a closed subspace of $L^{p,q}(R^{d+1})$.
\end{Lm}
\begin{Lm}\cite{ag}\label{L2.7}
If $\phi\in W(L^1)$ and $c\in \ell^1$, then the function $f=\sum \limits_{k\in Z^d}c_k\phi(x-k)$ belongs to $W(L^1)$ and
$$\|f\|_{W(L^1)}\leq C\|c\|_{\ell^1}\|\phi\|_{W(L^1)}.$$
\end{Lm}
\begin{Lm}\cite{lllz}\label{L2.8}
Suppose that $\phi\in W(L^1)(R^{d+1})$. Then for any $f\in L^{p,q}(R^{d+1})$, the sequence
$$c(k_1,k_2)=\int_R\int_{R^d}f(s,t)\phi(s-k_1,t-k_2)dsdt,\ k_1\in Z, k_2\in Z^d$$
belongs to $\ell^{p,q}(Z^{d+1})$ and $\|c\|_{\ell^{p,q}}\leq \|f\|_{L^{p,q}}\|\phi\|_{W(L^1)}$.
\end{Lm}
\begin{exam}
Suppose that $\varphi(x,y)$ satisfies the conditions in Lemma \ref{L2.6}. Then the function
\begin{equation}\label{r28}
K_1(x,y;s,t)=\sum \limits_{k_1\in Z}\sum \limits_{k_2\in Z^d}\varphi(x-k_1,y-k_2)\widetilde{\varphi}(s-k_1,t-k_2)
\end{equation}
satisfies \eqref{r1} and \eqref{r2}. Moreover, the shift-invariant subspace $V_{p,q}(\varphi)$ defined in \eqref{r27} is the range space of some idempotent integral operator with kernel $K_1(x,y;s,t)$ in \eqref{r28}. Furthermore, it is a reproducing kernel subspace of $L^{p,q}(R^{d+1})$.
\end{exam}
{\bf Proof}\ \ In fact, it is easy to verify that $$\|K_1\|_{\mathcal{W}}=\big\|\|K_1(x,y;s,t)\|_{\mathcal{W}^0_{y,t}}\big\|_{\mathcal{W}^0_{x,s}}\leq \|\varphi\|_{W(L^1)}\|\widetilde{\varphi}\|_{W(L^1)}<\infty.$$
Moreover, by the definition of the modulus of continuity,
\begin{align}\label{r31}
\omega_\delta(K_1)(x,y;s,t)&\leq \sum \limits_{k_1\in Z}\sum \limits_{k_2\in Z^d}|\varphi(x-k_1,y-k_2)|\omega_\delta(\widetilde{\varphi})(s-k_1,t-k_2)+\nonumber\\
&\ \ \ \sum \limits_{k_1\in Z}\sum \limits_{k_2\in Z^d}\omega_\delta(\varphi)(x-k_1,y-k_2)|\widetilde{\varphi}(s-k_1,t-k_2)|+\nonumber\\
&\ \ \ \sum \limits_{k_1\in Z}\sum \limits_{k_2\in Z^d}\omega_\delta(\varphi)(x-k_1,y-k_2)\omega_\delta(\widetilde{\varphi})(s-k_1,t-k_2).
\end{align}
Moreover, we know from \eqref{r30} that
\begin{equation}
\omega_\delta(\widetilde{\varphi})(x,y)\leq \sum \limits_{k_1\in Z}\sum \limits_{k_2\in Z^d}|b(k_1,k_2)|\omega_\delta(\varphi)(x-k_1,y-k_2).
\end{equation}
This together with Lemma \ref{L2.6} and Lemma \ref{L2.7} proves that $\lim \limits_{\delta\rightarrow 0}\|\omega_\delta(\widetilde{\varphi})\|_{W(L^1)}=0$. Finally, $\lim \limits_{\delta\rightarrow 0}\|\omega_\delta(K_1)\|_{\mathcal{W}}=0$ follows from
\begin{equation*}
\|\omega_\delta(K_1)\|_{\mathcal{W}}\leq \|\varphi\|_{W(L^1)}\|\omega_\delta(\widetilde{\varphi})\|_{W(L^1)}+\|\omega_\delta(\varphi)\|_{W(L^1)}\|\widetilde{\varphi}\|_{W(L^1)}+\|\omega_\delta(\varphi)\|_{W(L^1)}\|\omega_\delta(\widetilde{\varphi})\|_{W(L^1)}.
\end{equation*}
For $f\in L^{p,q}(R^{d+1})$, define
$$T_1f(x,y)=\int_R\int_{R^d}K_1(x,y;s,t)f(s,t)dsdt.$$
It is easy to verify that $T_1$ is an idempotent integral operator by the bi-orthogonality of $\varphi$ and $\widetilde{\varphi}$. Moreover, $V_{p,q}(\varphi)\subseteq T_1L^{p,q}$ follows from $T_1f(x,y)=f(x,y)$ for $f\in V_{p,q}(\varphi)$. For $f\in T_1L^{p,q}$,
\begin{align}
f(x,y)&=\int_R\int_{R^d}K_1(x,y;s,t)f(s,t)dsdt\nonumber\\
&=\sum \limits_{k_1\in Z}\sum \limits_{k_2\in Z^d}\varphi(x-k_1,y-k_2)\int_R\int_{R^d}f(s,t)\widetilde{\varphi}(s-k_1,t-k_2)dsdt\nonumber\\
&=:\sum \limits_{k_1\in Z}\sum \limits_{k_2\in Z^d}d(k_1,k_2)\varphi(x-k_1,y-k_2).
\end{align}
Then $f\in V_{p,q}(\varphi)$ due to $d\in \ell^{p,q}(Z^{d+1})$ by Lemma \ref{L2.8}. Therefore, $T_1L^{p,q}\subseteq V_{p,q}(\varphi)$. Finally, $T_1L^{p,q}= V_{p,q}(\varphi)$, which means that $V_{p,q}(\varphi)$ is just the range space of the idempotent integral operator $T_1$.

\section{Frame property }

\ \ \ \ In this section, we  show that the reproducing kernel subspace $V$ in \eqref{r26} has frames and has finite rate of innovation (\cite{s06}). Denote the standard action between functions $f\in L^{p,q}(R^{d+1})$ and $g\in L^{p',q'}(R^{d+1})$ by
$$\langle f,g\rangle=\int_R\int_{R^d}f(x,y)g(x,y)dxdy,$$
where $p',q'$ are the conjugate numbers of $p$ and $q$, respectively. We first introduce two definitions about $(p,q)$-frame and dual pair in mixed Lebesgue spaces.
\begin{Df}
Let $V$ be a Banach subspace of $L^{p,q}(R^{d+1})$. A family $\Phi=\{\psi_\gamma\}_{\gamma\in \Gamma}$ of functions in $L^{p',q'}(R^{d+1})$ is a $(p,q)$-frame for $V$, if there exist positive constants $A$ and $B$ such that
\begin{equation*}
A\|f\|_{L^{p,q}}\leq \|\{\langle f,\psi_\gamma\rangle\}_{\gamma\in \Gamma}\|_{\ell^{p,q}}\leq B\|f\|_{L^{p,q}},\ \forall f\in V.
\end{equation*}
\end{Df}
\begin{Df}
Let $1\leq p,q\leq \infty$, $V\subset L^{p,q}(R^{d+1})$ and $W\subset L^{p',q'}(R^{d+1})$. The $(p,q)$-frame $\widetilde{\Phi}=\{\widetilde{\phi}_\lambda\}_{\lambda\in \Lambda}\subset W$ for $V$ and the $(p',q')$-frame $\Phi=\{\phi_\lambda\}_{\lambda\in \Lambda}\subset V$ for $W$ form a dual pair if the following reconstruction formulae hold:
\begin{equation}\label{r23}
f=\sum \limits_{\lambda\in \Lambda}\langle f,\widetilde{\phi}_\lambda\rangle\phi_\lambda \ for \ all\ f\in V
\end{equation}
and
\begin{equation}\label{r24}
g=\sum \limits_{\lambda\in \Lambda}\langle g,\phi_\lambda\rangle\widetilde{\phi}_\lambda \ for \ all\ g\in W.
\end{equation}
\end{Df}
\begin{Tm}\label{T3.3}
Let $T$ be the idempotent integral operator on $L^{p,q}(R^{d+1})$ whose kernel $K$ satisfies \eqref{r1} and \eqref{r2}, $T^\ast$ be the adjoint of $T$, that is,
\begin{equation}
T^\ast g(x,y)=\int_R\int_{R^d}K(s,t;x,y)g(s,t)dsdt,\ g\in L^{p',q'}(R^{d+1}),
\end{equation}
and let $V$ and $V^\ast$ be the range spaces of $T$ on $L^{p,q}(R^{d+1})$ and $T^\ast$ on $L^{p',q'}(R^{d+1})$, respectively.
Then there exist a relatively-separately subset $\Lambda$, and two families $\Phi=\{\phi_\lambda\}_{\lambda\in \Lambda}\subset V$ and $\widetilde{\Phi}=\{\widetilde{\phi}_\lambda\}_{\lambda\in \Lambda}\subset V^\ast$ such that
 \begin{enumerate}
\item[(i)]  \ $\widetilde{\Phi}$ is a $(p,q)$-frame for $V$ and $\Phi$ is a $(p',q')$-frame for $V^\ast$;
\item[(ii)]  \  $\Phi$ and $\widetilde{\Phi}$ form a dual pair;
\item [(iii)] \ Both $V$ and $V^\ast$ are generated by $\Phi$ and $\widetilde{\Phi}$ respectively, in the sense that
\begin{equation}
V=\Big\{\sum \limits_{\lambda\in \Lambda}c(\lambda)\phi_{\lambda}:\ (c(\lambda))_{\lambda\in \Lambda}\in \ell^{p,q}(\Lambda)\Big\}
\end{equation}
and
\begin{equation}
V^\ast=\Big\{\sum \limits_{\lambda\in \Lambda}\widetilde{c}(\lambda)\widetilde{\phi}_{\lambda}:\ (\widetilde{c}(\lambda))_{\lambda\in \Lambda}\in \ell^{p',q'}(\Lambda)\Big\}.
\end{equation}
\end{enumerate}
\end{Tm}
{\bf Proof} Let $\delta$  be a sufficiently small positive number  such that
$$r_0(\delta)=:\max\Big\{\|K\|_\mathcal{W}\|\omega_{\sqrt{d+1}\delta}(K)\|_\mathcal{W}\Big(1+\frac{\|K\|_\mathcal{W}+\|\omega_{\sqrt{d+1}\delta}(K)\|_\mathcal{W}}{1-\|K\|_\mathcal{W}\|\omega_{\sqrt{d+1}\delta}(K)\|_\mathcal{W}}\Big), \|\omega_{\sqrt{d+1}\delta}(K)\|_\mathcal{W}\Big\}<1.$$
Define the operator $T_{\delta}$ by
\begin{equation}\label{r14}
T_{\delta}f(x,y)=\int_R\int_{R^d}K_{\delta}(x,y;s,t)f(s,t)dsdt,\ f\in L^{p,q}(R^{d+1}),
\end{equation}
where $$K_{\delta}(x,y;s,t)=\delta^{-1-d}\int_{-\delta/2}^{\delta/2}\int_{[-\delta/2,\delta/2]^d}\int_{-\delta/2}^{\delta/2}
\int_{[-\delta/2,\delta/2]^d}\sum \limits_{\lambda_1\in \delta Z}\sum \limits_{\lambda_2\in \delta Z^d}$$
$$K(x,y;\lambda_1+z_1,\lambda_2+z_2)K(\lambda_1+z_1',\lambda_2+z_2';s,t)dz_1dz_2dz_1'dz_2'.$$
Then it follows from \eqref{r12} that $T_{\delta}T=TT_{\delta}=T_{\delta}$ and
\begin{equation}\label{r15}
|K_{\delta}(x,y;s,t)-K(x,y;s,t)|\leq \int_R\int_{R^d}|K(x,y;z_1,z_2)|\omega_{\sqrt{d+1}\delta}(K)(z_1,z_2;s,t)dz_1dz_2.
\end{equation}
Therefore, for all $f\in L^{p,q}(R^{d+1})$, one has
\begin{align}\label{r16}
\|T_{\delta}f-Tf\|_{L^{p,q}}&\leq \|K_{\delta}-K\|_{\mathcal{W}}\|f\|_{L^{p,q}}\nonumber\\
&\leq \|K\|_{\mathcal{W}}\|\omega_{\sqrt{d+1}\delta}(K)\|_{\mathcal{W}}\|f\|_{L^{p,q}}\nonumber\\
&\leq r_0(\delta)\|f\|_{L^{p,q}}.
\end{align}
Then it follows from \eqref{r14}, \eqref{r15} and \eqref{r16} that the operator $$T_{\delta}^+:=T+\sum \limits_{n=1}^\infty(T-T_{\delta})^n$$ is a bounded integral operator which satisfies $T_{\delta}^+T_{\delta}=T_{\delta}T_{\delta}^+=T$. Let $K^+_{\delta}$ be the kernel of the operator $T_{\delta}^+$. Then
$$K^+_{\delta}(x,y;s,t)=K(x,y;s,t)+\big(K(x,y;s,t)-K_{\delta}(x,y;s,t)\big)+\sum \limits_{n=2}^\infty\int_R\int_{R^d}\cdots\int_R\int_{R^d}$$
$$\big(K(x,y;s_1,t_1)-K_{\delta}(x,y;s_1,t_1)\big)\big(K(s_1,t_1;s_2,t_2)-K_{\delta}(s_1,t_1;s_2,t_2)\big)$$$$\cdots
\big(K(s_{n-1},t_{n-1};s,t)-K_{\delta}(s_{n-1},t_{n-1};s,t)\big)ds_1dt_1\cdots ds_{n-1}dt_{n-1}.$$
Therefore, we have
\begin{align}\label{r21}
\|K^+_{\delta}\|_{\mathcal{W}}&\leq \|K\|_{\mathcal{W}}+\sum \limits_{n=1}^\infty\|K-K_{\delta}\|_{\mathcal{W}}^n\nonumber\\
&\leq \|K\|_{\mathcal{W}}+\frac{r_0(\delta)}{1-r_0(\delta)}<\infty.
\end{align}
For all $\lambda=(\lambda_1,\lambda_2)\in \delta Z\times\delta Z^d:=\Lambda$, define
\begin{equation*}
\phi_\lambda(x,y)=\delta^{-1/p-d/q}\int_R\int_{R^d}\int_{-\delta/2}^{\delta/2}\int_{[-\delta/2,\delta/2]^d}
K^+_{\delta}(x,y;z_1,z_2)\cdot
\end{equation*}
\begin{equation}
K(z_1,z_2;\lambda_1+z_1',\lambda_2+z_2')dz_1dz_2dz_1'dz_2'
\end{equation}
and
\begin{equation}
\widetilde{\phi}_{\lambda}(x,y)=\delta^{-1+1/p}\delta^{-d+d/q}\int_{-\delta/2}^{\delta/2}\int_{[-\delta/2,\delta/2]^d}
K(\lambda_1+z_1,\lambda_2+z_2;x,y)dz_1dz_2.
\end{equation}

\noindent(i)\ It follows from \eqref{r13} and the  Minkowski's inequality that
\begin{align*}
\|\widetilde{\phi}_{\lambda}\|_{L^{p',q'}}&\leq \delta^{-1+1/p}\delta^{-d+d/q}\int_{-\delta/2}^{\delta/2}\int_{[-\delta/2,\delta/2]^d}\|K(\lambda_1+z_1,\lambda_2+z_2;x,y)\|_{L^{p',q'}_{x,y}}dz_1dz_2\nonumber\\
&\leq\|K\|_{\mathcal{W}}^{\frac{1}{p'q'}} \Big(\|K\|_{\mathcal{W}}+\|\omega_{\sqrt{d+1}\delta}(K)\|_{\mathcal{W}}\Big)^{1-\frac{1}{p'q'}}.
\end{align*}
Therefore, $\widetilde{\phi}_{\lambda}\in L^{p',q'}(R^{d+1})$. Similarly,  $\phi_{\lambda}\in L^{p,q}(R^{d+1})$ follows from \eqref{r17} and
\begin{align*}
\|\phi_{\lambda}\|_{L^{p,q}}&\leq \delta^{-1/p}\delta^{-d/q}\int_{-\delta/2}^{\delta/2}\int_{[-\delta/2,\delta/2]^d}\Big\|\int_R\int_{R^d}K^+_{\delta}(x,y;z_1,z_2)
\\
& \cdot K(z_1,z_2;\lambda_1+z_1',\lambda_2+z_2')dz_1dz_2\Big\|_{L^{p,q}_{x,y}}dz'_1dz'_2\nonumber\\
&\leq \delta^{-1/p}\delta^{-d/q}\|K^+_{\delta}\|_{\mathcal{W}}\int_{-\delta/2}^{\delta/2}\int_{[-\delta/2,\delta/2]^d}
\|K(z_1,z_2;\lambda_1+z_1',\lambda_2+z_2')\|_{L^{p,q}_{z_1,z_2}}dz'_1dz'_2\\
&\leq \|K^+_{\delta}\|_{\mathcal{W}}\|K\|_{\mathcal{W}}^{\frac{1}{pq}} \Big(\|K\|_{\mathcal{W}}+\|\omega_{\sqrt{d+1}\delta}(K)\|_{\mathcal{W}}\Big)^{1-\frac{1}{pq}}.
\end{align*}
For any $x\in k_1\delta+[-\delta/2,\delta/2]$ and $y\in k_2\delta+[-\delta/2,\delta/2]^d$, one has
\begin{align}\label{r18}
|\langle f,\widetilde{\phi}_\lambda\rangle-\delta^{1/p}\delta^{d/q}f(x,y)|&\leq \delta^{-1+1/p}\delta^{-d+d/q}\int_R\int_{R^d}\int_{-\delta/2}^{\delta/2}\int_{[-\delta/2,\delta/2]^d}|K(k_1\delta+z_1,k_2\delta+z_2;s,t)\nonumber\\
&-K(x,y;s,t)||f(s,t)|dsdtdz_1dz_2\nonumber\\
&\leq \delta^{1/p}\delta^{d/q}\int_R\int_{R^d}\omega_{\sqrt{d+1}\delta}(K)(x,y;s,t)|f(s,t)|dsdt\nonumber\\
&=:\delta^{1/p}\delta^{d/q}F(x,y).
\end{align}
Define $$\alpha_{k_1}(x)=\chi_{k_1\delta+[-\delta/2,\delta/2]}(x),\ \beta_{k_2}(y)=\chi_{k_2\delta+[-\delta/2,\delta/2]^d}(y).$$
It follows from \eqref{r18} that
\begin{equation}\label{r19}
|\langle f,\widetilde{\phi}_\lambda\rangle|\alpha_{k_1}^{1/p}(x)\beta_{k_2}^{1/q}(y)\leq \delta^{1/p}\delta^{d/q}\alpha_{k_1}^{1/p}(x)\beta_{k_2}^{1/q}(y)(f(x,y)+F(x,y)).
\end{equation}
Taking the $\ell^q$-norm with respect to the variable $k_2\in Z^d$ on both sides of \eqref{r19} and then taking the $L^q$-norm with respect to the variable $y\in R^d$, one obtains
\begin{equation}\label{r20}
\alpha_{k_1}^{1/p}(x)\|\{\langle f,\widetilde{\phi}_\lambda\rangle\}_{k_2\in Z^d}\|_{\ell^q}\leq \delta^{1/p}\alpha_{k_1}^{1/p}(x)(\|f(x,\cdot)+F(x,\cdot)\|_{L^q}).
\end{equation}
Taking the $\ell^p$-norm for the variable $k_1\in Z$ on both sides of \eqref{r20} and then taking the $L^p$-norm for the variable $x\in R$, we have
$$\|\{\langle f,\widetilde{\phi}_\lambda\rangle\}_{\lambda\in \Lambda}\|_{\ell^{p,q}}\leq \|f\|_{L^{p,q}}+\|F\|_{L^{p,q}}\leq (1+\|\omega_{\sqrt{d+1}\delta}(K)\|_{\mathcal{W}})\|f\|_{L^{p,q}}.$$
Similarly, we can obtain $$\|\{\langle f,\widetilde{\phi}_\lambda\rangle\}_{\lambda\in \Lambda}\|_{\ell^{p,q}}\geq(1-\|\omega_{\sqrt{d+1}\delta}(K)\|_{\mathcal{W}})\|f\|_{L^{p,q}}$$
by the similar technique from the inequality $$\delta^{1/p}\delta^{d/q}\alpha_{k_1}^{1/p}(x)\beta_{k_2}^{1/q}(y)f(x,y)\leq |\langle f,\widetilde{\phi}_\lambda\rangle|\alpha_{k_1}^{1/p}(x)\beta_{k_2}^{1/q}(y)+\delta^{1/p}\delta^{d/q}\alpha_{k_1}^{1/p}(x)\beta_{k_2}^{1/q}(y)F(x,y).$$
Therefore, $\widetilde{\Phi}$ is a $(p,q)$-frame for $V$. $\Phi$ is a $(p',q')$-frame for $V^\ast$ can be similarly proved from
\begin{align}\label{r20}
&\ \ \ \ |\langle f,\phi_\lambda\rangle-\delta^{1-1/p}\delta^{d-d/q}f(x,y)|\nonumber\\
&=\delta^{-1/p}\delta^{-d/q}\int_R\int_{R^d}\Big[\int_R\int_{R^d}
\int_{-\delta/2}^{\delta/2}\int_{[-\delta/2,\delta/2]^d}\big|K^+_\delta(s,t;z_1,z_2)K(z_1,z_2;k_1\delta+z'_1,k_2\delta+z'_2)\nonumber\\
&\ \ \ \ -K(s,t;z_1,z_2)K(z_1,z_2;x,y)\big|dz_1dz_2dz'_1dz'_2\Big]|f(s,t)|dsdt \nonumber\\
&\leq \delta^{1-1/p}\delta^{d-d/q}\int_R\int_{R^d}\Big[\int_R\int_{R^d}|K^+_\delta(s,t;z_1,z_2)|\omega_{\sqrt{d+1}\delta}(K)(z_1,z_2;x,y)dz_1dz_2+\nonumber\\
&\ \ \ \ \int_R\int_{R^d}|K^+_\delta(s,t;z_1,z_2)-K(s,t;z_1,z_2)||K(z_1,z_2;x,y)|dz_1dz_2\Big]|f(s,t)|dsdt\nonumber\\
&=:\delta^{1-1/p}\delta^{d-d/q}(F_1(x,y)+F_2(x,y)).
\end{align}
Moreover, it follows from \eqref{r15}, \eqref{r16} and \eqref{r21} that
\begin{align*}
\|F_1+F_2\|_{L^{p',q'}}&\leq \big(\|K^+_\delta\|_\mathcal{W}\|\omega_{\sqrt{d+1}\delta}(K)\|_\mathcal{W}+\|K^+_\delta-K\|_\mathcal{W}\|K\|_{\mathcal{W}}\big)\|f\|_{L^{p',q'}}\\
&\leq \|K\|_\mathcal{W}\|\omega_{\sqrt{d+1}\delta}(K)\|_\mathcal{W}\Big(1+\frac{\|K\|_\mathcal{W}+\|\omega_{\sqrt{d+1}\delta}(K)\|_\mathcal{W}}{1-\|K\|_\mathcal{W}\|\omega_{\sqrt{d+1}\delta}(K)\|_\mathcal{W}}\Big)\|f\|_{L^{p',q'}}.
\end{align*}
(ii)\ It is easy to verify that $T^+_\delta=TT^+_\delta$, which means that
\begin{equation}\label{r22}
K^+_\delta(x,y;z_1,z_2)=\int_R\int_{R^d}K(x,y;s,t)K^+_\delta(s,t;z_1,z_2)dsdt.
\end{equation}
Moreover, we can prove $T\phi_\lambda=\phi_\lambda$ and $T^\ast\widetilde{\phi}_\lambda=\widetilde{\phi}_\lambda$ by \eqref{r22} and \eqref{r12}, respectively. Therefore, $\Phi\subset V$ and $\widetilde{\Phi}\subset V^\ast$. Furthermore, for all $f\in V$, one has
\begin{align*}
\sum \limits_{\lambda\in \Lambda}\langle f,\widetilde{\phi}_\lambda\rangle\phi_\lambda(x,y)&=
\delta^{-1-d}\int_R\int_{R^d}\int_R\int_{R^d}K^+_{\delta}(x,y;z_1,z_2)f(s,t)\Big[\sum \limits_{k_1\in Z}\sum \limits_{k_2\in Z^d}\int_{-\delta/2}^{\delta/2}\int_{[-\delta/2,\delta/2]^d}\\
&\ \ \ \ \int_{-\delta/2}^{\delta/2}\int_{[-\delta/2,\delta/2]^d}K(z_1,z_2;k_1\delta+z'_1,k_2\delta+z'_2)K(k_1\delta+u,k_2\delta+v;s,t)\\
&\ \ \ \ dz'_1dz'_2dudv\Big]dsdtdz_1dz_2\\
&=\int_R\int_{R^d}\int_R\int_{R^d}K^+_{\delta}(x,y;z_1,z_2)K_\delta(z_1,z_2;s,t)f(s,t)dz_1dz_2dsdt\\
&=T^+_\delta T_\delta f(x,y)=Tf(x,y)=f(x,y).
\end{align*}
Thus, \eqref{r23} is proved. \eqref{r24} can be proved similarly.

\noindent(iii)\ The result follows directly from (i) and (ii).

\section{Timing sampling and reconstruction}

\ \ \ \ In the final section, we  study a kind of semi-adaptive sampling and reconstruction of signals in the reproducing kernel subspace $V$.
Each device located at $y_j,j\in J$ first produces samples $f(x,y_j)$ in space domain, and then conducts the process of timing sampling in time domain by a time encoding machine (TEM). Since the sampling mechanism in time domains is adaptive to the signal, we call the overall sampling procedure for time-space signals as a semi-adaptive sampling pattern.

Two classes of TEMs have been considered in \cite{gv} for time signals in a $L^2$-shift invariant subspace. One is crossing TEM which relies on a test function and a comparator, sampling time instants are produced  when the signal crosses the test function.  The other is the Integrate-and-Fire time encoding machine which arises from the study of neurons \cite {Neu2002} and gives samples just like non-uniform average sampling, where the output is tuned to the variation of the integral of the signal.
The following mathematical definitions are borrowed from \cite{gv}, where the firing parameter $\alpha=0$. Here, we consider the more general case.

\begin{Df} \  A crossing TEM(C-TEM) with continuous test functions $\{\Phi_n\}$, outputs the sequence $\{t_n\}_{n \in \ZZ}$ such that
\begin{itemize}
\item [{(i)}]$\Phi_n$ may be recovered from the set $\{t_i: i\leq n-1\}$;
\item [{(ii)}]$f(t_n)=\Phi_n(t_n)$;
\item [{(iii)}]$f(t)\neq \Phi_n(t)$, for any $t\in (t_{n-1},t_n)$.
\end{itemize}
\end{Df}
\begin{Df} \  An Integrate-and-Fire TEM(IF-TEM) with  test functions $\{\Phi_n\}$ and firing parameter $\alpha\geq 0$, outputs the sequence $\{t_n\}_{n \in \ZZ}$ such that
\begin{itemize}
\item [{(i)}]$\Phi_n$ may be recovered from the set $\{t_i: i\leq n-1\}$;
\item [{(ii)}]$\int_{t_{n-1}}^{t_n}f(u)e^{\alpha(u-t_n)}du=\Phi_n(t_n)$;
\item [{(iii)}]$\int_{t_{n-1}}^{t}f(u)e^{\alpha(u-t)}du\neq \Phi_n(t)$, for any $t\in (t_{n-1},t_n)$.
\end{itemize}
\end{Df}
 If the output sequence $\{t_n\}_{n\in \ZZ}$ satisfies $t_{n+1}-t_n\leq \delta$ for any input signals, we call that such TEM is $\delta$-dense.  In this section, we always assume that $\lim \limits_{n\rightarrow \pm\infty}t_n=\pm \infty$, which corresponds to the models in \cite{gv,TEM04}.

 For time-space signals $f(x,y)$ in reproducing kernel subspace $V$, the samples produced by sampling devices with C-TEM are $\{f(x_i^{(j)},y_j):i\in Z, j\in J\}$. The sampling devices with IF-TEM give samples $\{\int_{x_{i}^{(j)}}^{x_{i+1}^{(j)}}f(u,y_j)e^{\alpha(u-x_{i+1}^{(j)})}du:i\in Z,j\in J\}$, which just like nonuniform average sampling for time variable.

 In the following, we  study how to reconstruct the signal $f(x,y)\in V$ from these kinds of samples.
Suppose that $U=\{u_j(y)\}_{j\in J}$ is a bounded uniform partition of unity associated with the covering $\{B(y_j,\delta')\}_{j\in J}$, which satisfies
\begin{itemize}
\item [{(i)}]\ $0\leq u_j(y)\leq 1$ for all $j\in J$ and $y\in R^d$;
\item [{(ii)}]\ $u_j(y)$ is supported in $B(y_j,\delta')$ for all $j\in J$;
\item [{(iii)}]\ $\sum \limits_{j\in J}u_j(y)\equiv 1$ for any $y\in R^d$.
\end{itemize}
Here, $\delta'$ is the gap of the relatively-separated set $\Gamma$. For example, we can take
$$u_j(y)=\frac{\chi_{B(y_j,\delta')}(y)}{\sum \limits_{k\in J}\chi_{B(y_{k},\delta')}(y)},\ j\in J.$$
Let $s_{n}^{(j)}:=\frac{x_{n-1}^{(j)}+x_{n}^{(j)}}{2}$, $j\in J$. For C-TEM, define the  operator $S$ by
 \begin{equation}
 Sf(x,y)=:\sum \limits_{j\in J}\sum \limits_{i\in Z}f(x_i^{(j)},y_j)\chi_{[s_i^{(j)},s_{i+1}^{(j)})}(x)u_j(y), \ f\in  V.
 \end{equation}

Based on this pre-reconstruction operator, the following theorem gives an iterative reconstruction algorithm with exponential convergence.

\begin{Tm}
Suppose that $K$ satisfies \eqref{r1} and \eqref{r2}, and that the C-TEM is $\delta$-dense. If
\begin{equation}
r_1=:\|K\|_{\mathcal{W}}\|\omega_{\sqrt{\delta^2+\delta'^2}}(K)\|_{\mathcal{W}}<1,
\end{equation}
then the iterative approximation algorithm
\begin{equation}
f_1=TSf,\ f_{n+1}=f_1+(I-TS)f_n
\end{equation}
satisfies $\|f-f_n\|_{L^{p,q}}\leq r_1^n\|f\|_{L^{p,q}}$ for all signals $f\in V$.
\end{Tm}
{\bf Proof}\ \ For any $x\in R$ and $j\in J$, there exists an $i_j\in Z$ such that $x\in [s^{(j)}_{i_j},s^{(j)}_{i_j+1})$. Then for any $f\in V$,
\begin{align}
|f(x,y)-Sf(x,y)|&=|f(x,y)-\sum \limits_{j\in J}f(x_{i_j}^{(j)},y_j)u_j(y)|\nonumber\\
&\leq \sum \limits_{j\in J}\int_R\int_{R^d}|K(x,y;s,t)-K(x_{i_j}^{(j)},y_j;s,t)|u_j(y)|f(s,t)|dsdt\nonumber\\
&\leq \int_R\int_{R^d}\omega_{\sqrt{\delta^2+\delta'^2}}(K)(x,y;s,t)|f(s,t)|dsdt.
\end{align}
Moreover, this together with Lemma \ref{L2.3} shows that for all $f\in V$,
\begin{align}
\|f-TSf\|_{L^{p,q}}&=\|Tf-TSf\|_{L^{p,q}}\nonumber\\
&\leq \|K\|_{\mathcal{W}}\|f-Sf\|_{L^{p,q}}\nonumber\\
&\leq \|K\|_{\mathcal{W}}\|\omega_{\sqrt{\delta^2+\delta'^2}}(K)\|_{\mathcal{W}}\|f\|_{L^{p,q}}\nonumber\\
&=r_1\|f\|_{L^{p,q}}.
\end{align}
Finally, we obtain that
$$\|f-f_n\|_{L^{p,q}}=\|(I-TS)^nf\|_{L^{p,q}}\leq r_1^n\|f\|_{L^{p,q}}$$
holds for all signals $f\in V$. The result is proved.


Based on the samples produced by IF-TEM, define the operator
\begin{equation}\label{r33}
Rf(x,y)=:\sum \limits_{j\in J}\sum \limits_{i\in Z}\int_{x_{i}^{(j)}}^{x_{i+1}^{(j)}}f(u,y_j)e^{\alpha(u-x_{i+1}^{(j)})}duK(x,y;s_{i+1}^{(j)},y_j)\|u_j\|_{L^1},\ f\in V,
\end{equation}
which  plays an important role in the corresponding iterative algorithm in Theorem \ref{T2}.
\begin{Lm}
The operator $R$ defined in \eqref{r33} is a bounded operator from $V$ to $V$.
\end{Lm}
{\bf Proof}\ \ Since $K(x,y;s_{i+1}^{(j)},y_j)\in V$ for all $i\in Z, j\in J$,  $Rf\in V$ for any $f\in V$. Note that
\begin{align*}
|Rf(x,y)|&\leq\sum \limits_{j\in J}\sum \limits_{i\in Z}\int_{x_{i}^{(j)}}^{x_{i+1}^{(j)}}\int_{R^d}\int_R\int_{R^d}|K(u,y_j;s,t)K(x,y;s_{i+1}^{(j)},y_j)||f(s,t)|u_j(v)dudvdsdt\\
&\leq \int_R\int_{R^d}\int_R\int_{R^d}\Big(|K(x,y;u,v)|+\omega_{\sqrt{\delta^2+\delta'^2}}(K)(x,y;u,v)\Big)\cdot\\
&\ \ \Big(|K(u,v;s,t)|+\omega_{\sqrt{\delta^2+\delta'^2}}(K)(u,v;s,t)\Big)|f(s,t)|dudvdsdt
\end{align*} holds for all $f\in V$.
Based on this estimation, we  obtain
$$\|Rf\|_{L^{p,q}}\leq \Big(\|K\|_{\mathcal{W}}+\|\omega_{\sqrt{\delta^2+\delta'^2}}(K)\|_{\mathcal{W}}\Big)^2\|f\|_{L^{p,q}}.$$

\begin{Tm}\label{T2}
Suppose that $K$ satisfies \eqref{r1} and \eqref{r2}, and that the IF-TEM is $\delta$-dense. If
\begin{align}
r_2&=\|K\|_{\mathcal{W}}\Big[\|\omega_{\sqrt{\delta^2+\delta'^2}}(K)\|_{\mathcal{W}}\big(2\|K\|_{\mathcal{W}}+
\|\omega_{\sqrt{\delta^2+\delta'^2}}(K)\|_{\mathcal{W}}\big)+\nonumber\\
&\ \ \ (1-e^{-\alpha\delta})\big(\|K\|_{\mathcal{W}}+
\|\omega_{\sqrt{\delta^2+\delta'^2}}(K)\|_{\mathcal{W}}\big)^2\Big]<1,
\end{align}
then the iterative approximation algorithm
\begin{equation}
f_1=Rf,\ f_{n+1}=f_1+(I-R)f_n
\end{equation}
satisfies $\|f-f_n\|_{L^{p,q}}\leq r_2^n\|f\|_{L^{p,q}}$ for all signals $f\in V$.
\end{Tm}
{\bf Proof}\ \ For any $g\in L^{p',q'}(R^{d+1})$, define
$$\widetilde{S}g(x,y)=\sum \limits_{j\in J}\sum \limits_{i\in Z}g(s_{i+1}^{(j)},y_j)\|u_j\|_{L^1}\int_{x_{i}^{(j)}}^{x_{i+1}^{(j)}}e^{\alpha(u-x_{i+1}^{(j)})}K(u,y_j;x,y)du.$$
For any $f\in V$ and $g\in V^\ast$, one has
\begin{align*}
\langle g,Rf\rangle&=\sum \limits_{j\in J}\sum \limits_{i\in Z}\int_R\int_{R^d}f(u,y_j)u_j(v)e^{\alpha(u-x_{i+1}^{(j)})}g(s_{i+1}^{(j)},y_j)\chi_{[x_i^{(j)},x_{i+1}^{(j)})}(u)dudv\\
&=\int_R\int_{R^d}f(s,t)\Big[\sum \limits_{j\in J}\sum \limits_{i\in Z}g(s_{i+1}^{(j)},y_j)\int_R\int_{R^d}u_j(v)e^{\alpha(u-x_{i+1}^{(j)})}
\\
&\ \ \chi_{[x_i^{j},x_{i+1}^{j})}(u)K(u,y_j;s,t)dudv\Big]dsdt\\
&=\langle f,\widetilde{S}g\rangle=\langle f,T^\ast\widetilde{S}g\rangle.
\end{align*}
Therefore, $R^\ast=T^\ast\widetilde{S}$ on $V$ and $V^\ast$. Moreover, for any $g\in V^\ast$, we have
\begin{align}\label{r34}
|g(x,y)-\widetilde{S}g(x,y)|&\leq \int_R\int_{R^d}\Big[\sum \limits_{j\in J}\sum \limits_{i\in Z}\int_{x_{i}^{(j)}}^{x_{i+1}^{(j)}}\int_{R^d}\big|K(s,t;u,v)K(u,v;x,y)-\nonumber\\
&\ \ e^{\alpha(u-x_{i+1}^{(j)})}K(s,t;s_{i+1}^{(j)},y_j)K(u,y_j;x,y)\big|u_j(v)dudv\Big]|g(s,t)|dsdt\nonumber\\
&\leq\int_R\int_{R^d}\Big[\sum \limits_{j\in J}\sum \limits_{i\in Z}\int_{x_{i}^{(j)}}^{x_{i+1}^{(j)}}\int_{R^d}\big|K(s,t;u,v)K(u,v;x,y)-\nonumber\\
&\ \ \ K(s,t;s_{i+1}^{(j)},y_j)K(u,y_j;x,y)\big|u_j(v)dudv\Big]|g(s,t)|dsdt+\nonumber\\
&\ \ \ (1-e^{-\alpha\delta})\int_R\int_{R^d}\Big[\sum \limits_{j\in J}\sum \limits_{i\in Z}\int_{x_{i}^{(j)}}^{x_{i+1}^{(j)}}\int_{R^d}\big|K(s,t;s_{i+1}^{(j)},y_j)K(u,y_j;x,y)\big|\nonumber\\
&\ \ \ u_j(v)dudv\Big]|g(s,t)|dsdt\nonumber\\
&=:I+(1-e^{-\alpha\delta})II.
\end{align}
Furthermore, we have
\begin{equation*}
\|I\|_{L^{p',q'}}\leq \|\omega_{\sqrt{\delta^2+\delta'^2}}(K)\|_{\mathcal{W}}\big(2\|K\|_{\mathcal{W}}+\|\omega_{\sqrt{\delta^2+\delta'^2}}(K)\|_{\mathcal{W}}\big)\|g\|_{L^{p',q'}}
\end{equation*}
and
\begin{equation*}
\|II\|_{L^{p',q'}}\leq \big(\|K\|_{\mathcal{W}}+\|\omega_{\sqrt{\delta^2+\delta'^2}}(K)\|_{\mathcal{W}}\big)^2\|g\|_{L^{p',q'}}.
\end{equation*}
These together with \eqref{r34} show that
\begin{equation}\label{r35}
\|g-\widetilde{S}g\|_{L^{p',q'}}\leq \frac{r_2}{\|K\|_{\mathcal{W}}}\|g\|_{L^{p',q'}},\ g\in V^\ast.
\end{equation}
Moreover, it follows from \eqref{r35} that
\begin{equation}\label{r36}
\|I-R\|_{V}=\|I-T\widetilde{S}\|_{V^\ast}\leq \|K\|_{\mathcal{W}}\|I-\widetilde{S}\|_{V^\ast}\leq r_2.
\end{equation}
Finally, $\|f-f_n\|_{L^{p,q}}=\|(I-R)^nf\|_{L^{p,q}}\leq r_2^n\|f\|_{L^{p,q}}$ for all signals $f\in V$.\\

\noindent \textbf{Acknowledgement }\ \ The project is partially
supported by the National Natural Science  Foundation of China (Nos. 11661024, 11525104, 11531013, 11671107) and the  Guangxi Natural Science Foundation
(No. 2017GXNSFAA198194), Guangxi Key
Laboratory of Cryptography and Information Security (No. GCIS201614), Guangxi
Colleges and Universities Key Laboratory of Data Analysis and
Computation.

 \end{document}